# Experimental demonstration of a multiphysics cloak: manipulating heat flux and electric current simultaneously

Yungui Ma[1,*], Yichao Liu[1], Muhammad Raza[1], Yudong Wang[1], and Sailing He[1,2,*]

*[1]State Key Lab of Modern Optical Instrumentation, Centre for Optical and Electromagnetic Research, Department of Optical Engineering, Zhejiang University, Hangzhou 310058, China*
*[2]Department of Electromagnetic Engineering, School of Electrical Engineering, Royal Institute of Technology, S-100 44 Stockholm, Sweden*

 *Correspondence should be addressed to: yungui@zju.edu.cn or sailing@kth.se

In past years, triggered by their successful realizations in electromagnetics, invisible cloaks have experienced rapid development and have been widely pursued in many different fields, though so far only for a single physical system. In this letter we made an unprecedented experimental attempt to show a multidisciplinary framework designed on the basis of two different physical equations. The proposed structure has the exceptional capability to simultaneously control two different physical phenomena according to the predetermined evolution scenarios. As a proof of concept, we implemented an electric-thermal bifunctional device that can guide both electric current and heat flux "across" a strong 'scatter' (air cavity) and restore their original diffusion directions as if nothing exists along the paths, thus rending dual cloaking effects for objects placed inside the cavity. This bifunctional cloaking performance is also numerically verified for a point-source nonuniform excitation. Our results and the fabrication technique presented here will help broaden the current research scope for multiple disciplines and may pave a prominent way to manipulate multiple flows and create new functional devices, e.g., for on-chip applications.

The state and evolution of one physical phenomenon (wave or flux) are fundamentally determined by their manner of interaction with surrounding matters abiding different physical theorems or laws, mathematically described by different characteristic equations, such as Maxwell's equations for electromagnetic (EM) waves or conduction equation for current/heat flow. According to Einstein's theory of general relativity, these differential equations representing basic physical laws will experience formality invariance (but with their parametrical coefficients changed) if transformed among different coordinate systems [1,2]. The covariance of free-space Maxwell's equations successfully explains the constant nature of light velocity in different inertia systems [3] and also led to the important conclusion that wave behaviors in curved empty geometries are equivalent to electromagnetism in a medium [4,5]. In 2006, Pendry and Leonhardt developed this important conception on coordinate mapping with their independent proposals that light could be bent to realize invisible cloaking [6,7], which was quickly verified in the experiment [8]. This scientific milestone on cloaking has triggered enormous research enthusiasm on transformation optics (TO) and their technological applications associated with the development of metamaterials [9,10], which has greatly contributed to the advancement of modern electromagnetism [11]. In past years people working in this field were mostly endeavored to improve the practicality of EM cloaks with various versions of designing algorithms and have successfully demonstrated the cloaking effect covering nearly the whole frequency range from static to visible light [12-21]. In addition to EM cloaks, other instrumental EM devices with functionalities deemed impossible before have been deeply explored using the transformations [22-24]. Meanwhile, the same technique has been theoretically generalized and applied to different physical equations. Accordingly, functional devices like invisible cloaks for various systems such as acoustic, matter wave, or heat flux, have been proposed which vastly enriched the research scope of the field and allowed unprecedented ways to control and manipulate these waves or flux in the pursuit of new physics and technological applications [25-33].

Nevertheless, the strong potential of the transformation technique is a challenging practice. The freedom of spatial manipulation gives rise to very complicated materials consisting of inhomogeneous tensor components often with extreme parametric requirements. For example, an EM cloak has to take into account both electric and magnetic polarizations, either of which has a maximum number of nine required dynamic elements [6], while a transient thermal cloak must deal with anisotropic thermal conductivities, thermal capacitance and density simultaneously [34]. Most of the previous work in this field struggled to satisfy the requirements of one set of physical equations that only controlled and realized one physical phenomenon. Numerous successful applications of this technique have been witnessed in different physical domains, in particular for invisible cloaking, which has been thoroughly explored in almost all applicable sub-fields. Scientifically, it will be very interesting to check whether it is possible to couple different equations in one entity and realize more than one physical phenomenon with a single device, i.e., achieving multiphysics through single or multiple transformations. The multiphysics capability will definitely help broaden the scope of the current field as well as provide unique opportunities for new applications. If not bothered by the fabrication issue, one can easily configure such a mixed model by applying transformations to different equations that will lead to a system of materials consisting of multiple sets of complicated physical parameters. At a first glance it seems very infeasible to build such a complex device according to the effective medium theory due to the lack of enough variables (e.g., ingredient species, topology, volume ratio, etc.) to solve these coupled equations. Untill now there is no experimental report on such a possibility except for two theoretical papers that focused on the composite material approach to realize a thermal-electric bifunctional transformed device [35,36]. In this letter we make such a an unprecedented experimental attempt to implement a multidisciplinary framework that has the capability to respond to two kinds of different physical phenomena and evolves according to the predesigned scenarios. As a proof of concept we start with the combination of the two fundamental conduction flows, i.e., electric current and heat flux, which share the same material homogenization algorithm, and fabricate a silicon-based artificial device employing a standard microfabrication technique. Experimentally we show that both electric current and heat flux can be elegantly manipulated to produce a bifunctional cloaking effect, which is robust with respect to different excitations and cloaking topologies. Our results indicate the practical possibility to implement multidisciplinary functional devices via relatively available technical schemes, which may pave a new way to create chip-oriented devices or technologies.

For the simplicity of implementation we concentrate on static conduction equations in two dimensions (2D). The methodology to realize the functionality presented here can be equally extended to three dimensions (3D) with modified constituent equations. It is noted that bifunctional electric-thermal devices and their synthesis methods have been theoretically explored in the previous literature employing the transformation technique [35,36]. To implement the composite material approaches presented by them, one needs to overcome some practical issues such as how to precisely fabricate and orderly mix different species of inclusions in a micro scale and effectively suppress the interfacial impedance. Here we take another approach to pursue the bifunctional property by including the material approach previously developed to achieve a DC magnetic cloak [37], and more recently for a 2D/3D thermal cloak [32,33]. In principle, this is a bilayer shell system with one layer to attract the field lines and the other to expel the lines and works at the static or quasi-static condition assuming under homogeneous external excitations. The constituent media should have a large contrast in their properties. In our case, we assume one of them is air hole and the other is a solid that has high electric and thermal conductivities ($\sigma$ and $\kappa$). As shown in the left of Fig. 1(a), the air cavity (black) is regarded as a strong scatter that absolutely expels the current/heat flow while the high-indexed solid (orange) in the middle acts oppositely to concentrate them. In an optimal combination of these two materials, as schematically shown in the right of Fig. 1(a), the disturbance to the homogenous external exciting fields can be utterly balanced out by their counteractions and thus render a cloaking effect for any object placed inside the cavity.

The material parameters with given cloaking topologies can be obtained from the solutions of conduction equations satisfying the above conditions. For the electric and the thermal Laplace conduction equations explored here, i.e., $\nabla \cdot (\sigma \nabla V) = 0$ and $\nabla \cdot (\kappa \nabla T) = 0$, their general solutions $\Psi$ in the cylindrical coordinate system ($r$, $\theta$) can be expressed as the summation of different orders of eigen functions, i.e., $\Psi = \sum_{n=1}^{\infty} \left( A_n r^n + B_n r^{-n-1} \right) P_n(\cos\theta)$ where $A_n$ and $B_n$ are the coefficients of the n$^{th}$ order Legendre polynomial $P_n$. For a source-free system, applying standard boundary continuity conditions and assuming homogeneous exciting fields [37], we can solve the unknown coefficients and obtain the analytic expressions for the conductivities of the second layer (orange shell)

$$\sigma_2 = \frac{R_2^2 + R_1^2}{R_2^2 - R_1^2} \sigma_b, \quad (1a)$$

$$\kappa_2 = \frac{R_2^2 + R_1^2}{R_2^2 - R_1^2} \kappa_b, \quad (1b)$$

where $R_1$ and $R_2$ are the inner and external radii of the shell, and $\sigma_b$ and $\kappa_b$ are the electric and thermal conductivities of the background. With a given background the conduction parameters for the cloaking shell are decided by the ratio of the external and inner radii $R_2/R_1$, as shown in Fig. 1(b). To implement this device we only need to satisfy the relation $\sigma_2/\sigma_b = \kappa_2/\kappa_b$, which can be realized by two different materials with equal $\sigma/\kappa$ ratios. Conventional metallic or dielectric materials will not be good options for this purpose due to their extreme electric properties [29-33]. Here we use a typical semiconductor material, low-density doped n-type silicon with moderate electric and thermal properties ($\sigma$ = 1 $\Omega^{-1}$cm$^{-1}$ and $\kappa$ = 149 W·m$^{-1}$K$^{-1}$ at room temperature), as the external shell of our cloak. For the background the same silicon material is utilized, but with periodically perforated air holes are utilized. These holes are filled with polydimethylsiloxane (PDMS) that has a thermal conductivity about 0.15 W·m$^{-1}$K$^{-1}$. According to the effective medium theory and the fact that PDMS fillings have negligible conductivities compared to silicon, this composite background will guarantee a matched $\sigma/\kappa$ ratio with the cloak shell. As shown in Fig. 1(b), the volume ratio $\Gamma$ of silicon in the background can be modulated to acquire the desired thermal and electric properties in terms of Eqs. 1(a) and 1(b) [35,36]. Note that our thermal measurement is carried out in a relatively narrow temperature window from 30 to 50 °C that corresponds to a variation of $\sigma$ or $\kappa$ less then 10%. Simulation indicates that this small variation has little influence on device performance. Figure 1(c) shows a photograph of our 0.5-mm-thick rectangular sample implemented with the parameters $R_1$= 5 mm, $R_2$ = 10 mm, $p$ = 2 mm and $d$ = 0.8 mm, where $p$ and $d$ are the period and width of squared holes in the background. These parameters are computer optimized and correspond to a background of $\sigma_{eff}$ = 0.6 $\Omega^{-1}\cdot cm^{-1}$ and $\kappa_{eff}$ = 89.3 W·m$^{-1}$K$^{-1}$. For measurement our electric and thermal sources are applied on the left edge of the sample and the sinks on the right edge.

Figures 2(a) and 2(b) give the simulated spatial potential distributions without and with a cloak design, respectively. The software COMSOL MULPHYSICS has been employed in our work to simulate the models. In the electric domain, the left and right boundaries are defined with potential 1 V and 0, respectively, and the top and bottom boundaries with electric insulation conditions (i.e., there is no current passing through these boundaries). The dashed black lines depict the external profile of the cloak and the white solid lines represent the potential contours that are orthogonal to the current flow trajectories. As shown in Fig. 2(a) for the controlling sample, the embedded air cavity causes a strong disturbance on the current flow and forms a field 'shadow' after it. This disturbing effect is completely suppressed by applying a 5-mm-thick cloaking shell, as evidenced by the straight potential contour lines right after the cloak as shown in Fig. 2(b). Our measurement setup is schematically shown in Fig. 2(c) where a pair of DC micro probers are used to map the spatial voltage signals across the whole sample surface. External 1-V electric bias is uniformly applied on the sample through two 200-nm-thick aluminum strips coated on the sample edges. Local potential is accurately measured through an array of Al microelectrodes at a spatial resolution equal to the lattice constant of the perforated background. The measured 2D distribution is given in Fig. 2(d), where spatial interpolation has been made to provide enough data points. From the voltage contours, it is seen that the electric currents are successfully guided around the air cavity and their straight trajectories are restored after passing the cloaking shell, which is in good agreement with the simulation. The wave-like disturbance of the contour lines manifests the periodic feature of the background material, which can be minimized by employing smaller unit cells at the same volume ratio for silicon.

Figures 3(a) and 3(b) show the simulated static temperature profiles for the samples without and with a cloaking shell, respectively. In the numerical model, the left and right boundaries are set at 50 and 30 °C, respectively, and the top/bottom boundary satisfies the thermal insulation condition. The white solid lines denote the isothermal lines that are orthogonal to the heat flux trajectories. In Fig. 3(a), the air cavity alone expels the heat flux and causes a temperature 'shadow' on its right side. With a desired cloaking shell, as shown in Fig. 3(b), this disturbance on the heat flux is totally diminished, as evidenced by the straight isothermal lines. Objects inside the cavity in principle will have less effect on the external temperature field and thus will be thermally cloaked. In our discussion we have neglected the influence of thermal convection and radiation on the heat transfer and only considered the conduction process. This assumption is valid for a relatively narrow temperature window as studied here. Figure 3(c) gives the schematic of our heat characterization setup. Two water tanks of constant temperatures 50 °C and 30 °C are used as the heat source and sink, respectively, and are connected to the sample through 0.5-mm-thick copper foils. An infrared camera placed on top of the sample captures the temperature field through the thermal emission effect. After immersing the sample in the tanks for 300 s, the whole system nearly approaches a thermally equilibrium state corresponding to a stable temperature profile. As shown in Fig. 3(d), the measured stable temperature profile shows an obvious thermal cloaking effect by minimizing the 'scattering' of the air cavity on the heat flux, which is in good agreement with the numerical prediction in Fig. 3(b). The slight difference may be ascribed to the existences of the parameter drift due to temperature variation and other heat exhausting channels that are not considered in simulation.

In the above we have experimentally verified the bifunctinal cloaking performance of our bilayer device under uniform field excitations. For practical application the incident field can be excited in any form. In this part we numerically examine the cloaking effect of our device impinged with nonuniform fields. Figures 4(a) and 4(d) plot the voltage and temperature distributions under point-source excitation for two different cloaks with $R_2/R_1$ = 2 and 1.02, respectively. The middle and right columns in Fig. 4 show the contrast effects from the sample without the cloaking shell and the empty background, respectively. The field distributions in these figures are normalized by the maximum value and do not differ between the electric and thermal scenarios. Comparatively, it is seen that our cloaking device maintains the desired current/flux manipulation capabilities even for the point-source excitation, although it is originally designed for a uniform field excitation. This characteristic property is robust and insensitive to the thickness of the cloaking shell, offering another valuable freedom for practical implementation. A similar effect has been previously found for DC magnetic field cloaks [37]. These features correlatewith the near-field disturbance nature for the conduction flows controlled by Laplace equations. A similar effect is anticipated for EM waves but only at the quasi-static limit where multilayer shells can be configured to hide a metal wire by suppressing the scattering from the lowest order modes [38].

In conclusion, we have provided the first experimental demonstration of a multidisciplinary artificial device that can work on both electric current and heat flux according to predetermined scenarios. For static application, our device exhibits good cloaking performance at the excitations of various source fields. We have also found that the present simple device has a reasonably good transient thermal response (not given here), particularly for a cloak with a large $R_2/R_1$ ratio. The capability to simultaneously deal with both electric current and heat flux can provide unique advantages to manipulate their physical phenomena for some specific applications and may be employed in designing high-performance solar cell systems [39] or on-chip component integration where current and heat are both critically involved [40]. The successful application of silicon material thus has its practical potential.

The methodologies presented here can be regarded as a universal case for a static physical phenomenon and in principle can be extended to realize the bifunctional response for other combinations of double or multiple systems controlled by similar Laplace equations. For this purpose, multiple functional or bifunctional shells dealing with different systems separately in space may be introduced to further improve the fabrication, which promises the integration of various functionalities (e.g. cloaking and concentration) in one device. Our current results clearly show such a possibility to achieve a multi-functional device with engineered effective media, which may afford new opportunities to conceive unprecedented technological applications.

**Acknowledgements**

The authors are grateful to the partial supports from NSFCs 61271085, 60990322 and 91130004, the National High Technology Research and Development Program (863 Program) of China (No. 2012AA030402), the Program of Zhejiang Leading Team of Science and Technology Innovation, NCET, MOE SRFDP of China, and Swedish VR grant (# 621-2011-4620) and SOARD and the support by the Fundamental Research Funds for the Central Universities.

**Figure 1**

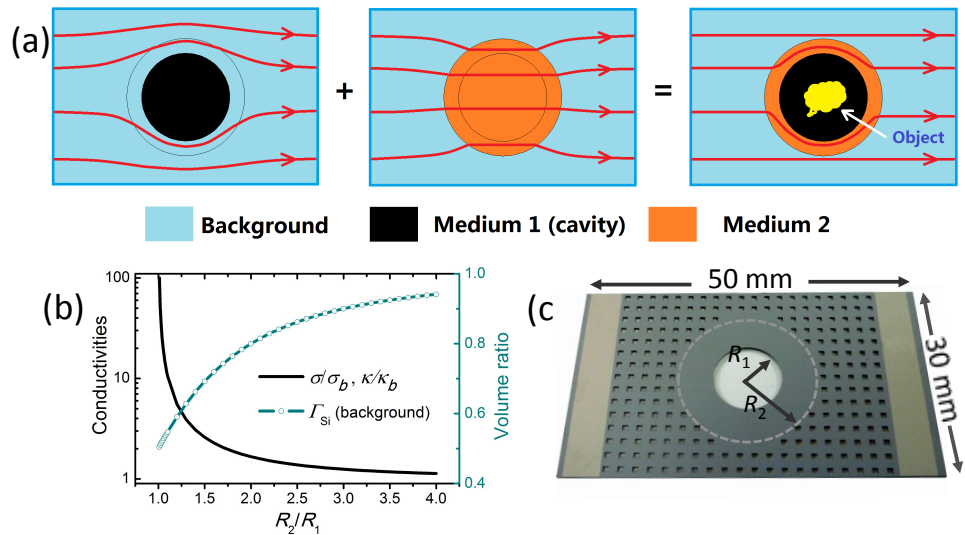

**Figure 1** (Color online) Design of bifunctional cloak. (a) From the left to right: an empty cavity (black), a medium (orange) with high conductive parameters and a bilayer cloak (black + orange) in the background. The red lines represent the conduction flow (i.e., heat or current). (b) Relative electric and thermal conductivities of the cloak and the silicon volume ratio in the background as a function of the radius ratio $R_2/R_1$ of the cloaking shell. (c) A photograph of fabricated sample device. $R_1$ ($R_2$) denotes the inner (outer) radius of the cloaking shell.

**Figure 2**

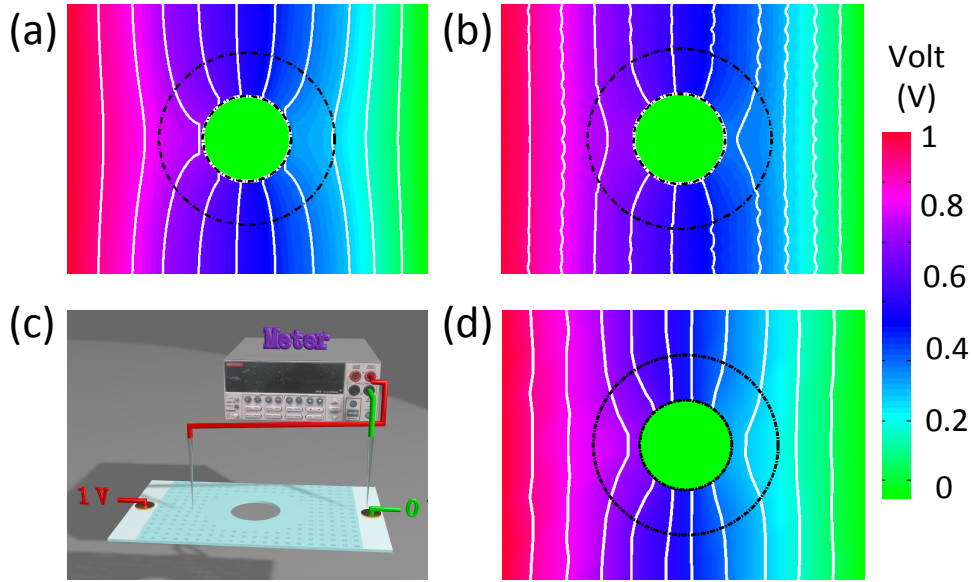

**Figure 2** (Color online) Electric current cloaking. Numerical electric potential profiles of (a) the controlling sample without a cloaking shell and (b) the sample with a cloaking shell. (c) Electric measurement setup. One end of the sample is biased at 1 V and the other is grounded. The spatial potential distribution is mapped by a voltage meter through a pair of micro probers. (d) Measured voltage distribution of the sample. In (a), (b) and (c), the black dashed lines describe the cloaking shell and the white solid lines denote the potential contours that are orthogonal to the current lines.

**Figure 3**

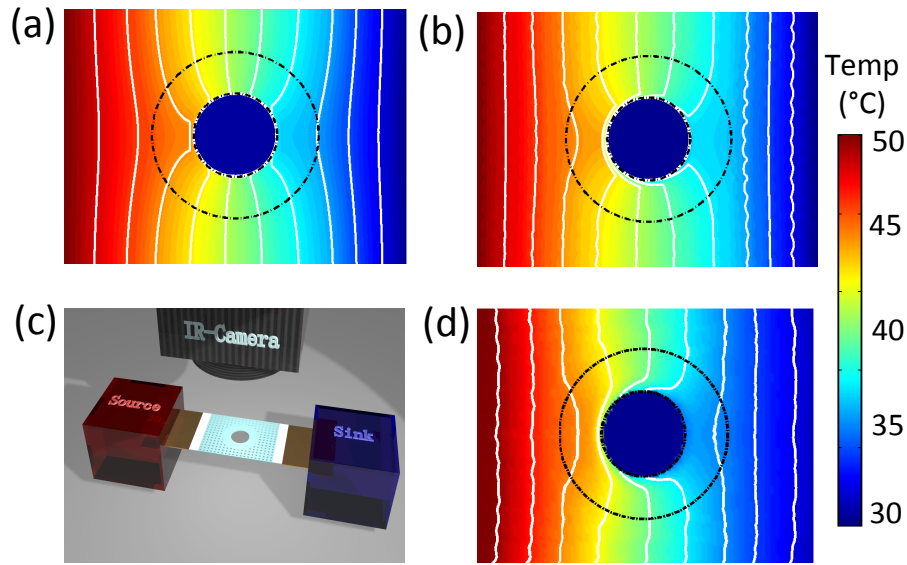

**Figure 3** (Color online) Heat flux cloaking. Simulated temperature profiles for the samples (a) without and (b) with a cloaking shell. The white solid lines denote the iso-thermal temperature curves that are orthogonal to the heat flux trajectories. (c) Heat-transfer measurement setup. Two constant-temperature water tanks are used as the source (50 °C) and sink (30 °C). An infrared camera through thermal radiation emission captures the temperature field. (d) Measured temperature profile of the sample at the thermal equilibrium state.

**Figure 4**

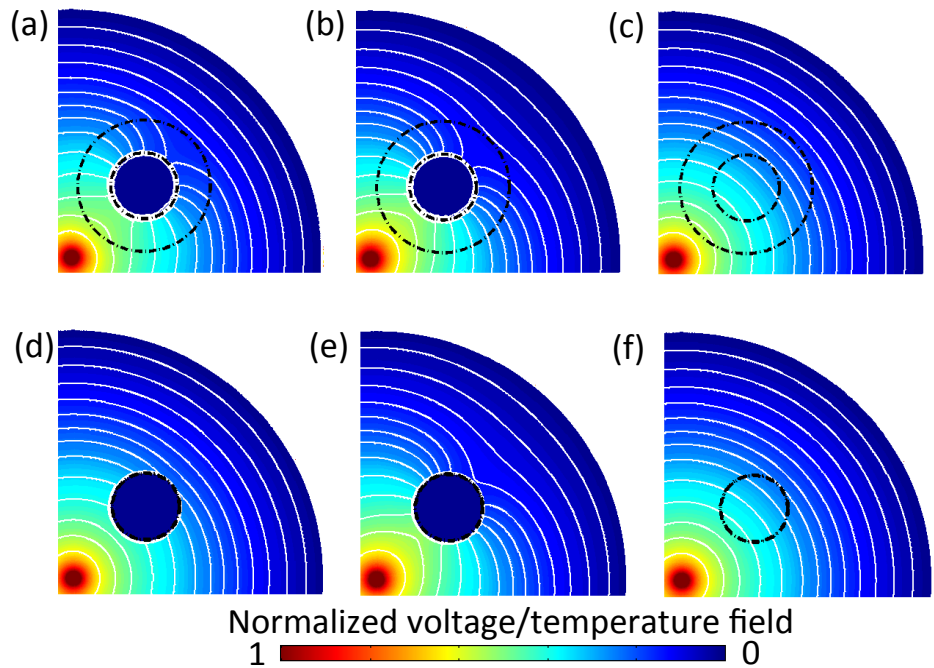

Figure 4 (Color online) Simulated bifunctional cloaking performance in a point-source external field. As denoted by the black dashed lines, figures in the top and bottom arrows correspond to the samples of $R_2/R_1$ = 2 and 1.02 (with a fixed $R_1$), respectively. (a)/(d), (b)/(e) and (c)/(f) correspond to the cases for a cloak, a cavity and the empty background, respectively. These results equally apply to both electric current and heat flux in our discussions.